**Modeling the self-energy and wavefunction relaxation in the orbitals**

B Huang

*Department of Physics and Materials Science, City University of Hong Kong, Kowloon, Hong Kong SAR, China*
*Present Address: Department of Applied Biology and Chemical Technology, The Hong Kong Polytechnic University, Hung Hom, Kowloon, Hong Kong SAR, China*

Email: bhuang@polyu.edu.hk

**Abstract**

The strong boundary normalized condition of wavefunction for fully occupied semicore 3d orbitals leads the linear response DFT+U on such metal oxide to have an insurmountable obstacle in Hubbard U determination. We treated the orbital self-energy and orbital relaxation as components of eigenvalues with respective orbital occupation number that follows the Fermi-Dirac distribution. By self-consistently solving the second partial deviation of total energy based on the most simple local density formalism with Hubbard U correction, we found the local density exchange-correlation potential functional can only give a minimized residue of the self-energy and orbital relaxation on the focus orbital if the Janak theorem maintained. Such residue turns to well counteracted in the fully occupied orbitals and non-zero the partially occupied orbitals. With keeping the validation of Janak theorem on localized orbitals, the self-consistent cycle by local density functional with Hubbard U correction cannot find out a set of orbital occupation that simultaneously offsets the orbital self-energy and relaxations in the empty and partially filled shell, but returns a unique set of the occupation for fully occupied shell. The band gap calculations on fully occupied orbital based compounds are thus improved and the relaxed lattices are also shown based on minimization of the self-energy error, which shows a possible route for accurate excited state studies.

## I. Introduction

The DFT+ (Hubbard U[1]) method on fully occupied semicore orbitals in $Cu_2O$ and $ZnO$ or other similar metal oxide materials has suffered large errors of underestimated band gaps (e.g. 0.73 eV in GGA compared to 3.44 eV from experimental results). If we look at the valence orbitals of $Zn^{2+}$ and $O^{2-}$ (if assume completed charge transfer of two valences) are both fully occupied shell such as $3d^{10}$ and $2p^6$ for cations and anions respectively. A significant method called linear response developed by the group of Cococcioni[2] has hereby faced an obstacle, which originally induced potential perturbation on specific partially filled orbital linear to charge occupancy to solve Hubbard U self-consistently out following Janak theorem[3]. As is well known, this method has been successful in the system with partially filled orbitals[4, 5].

It merged two difficulties: One is physical, the electron wavefunction of $3d^{10}$ are constrained with strict boundary conditions, such orbital-based potential perturbation to vary occupancies is nearly insurmountable; another is mathematical, the perturbation becomes extremely small if a small Lagrange multiplier is used to perturb the fully filled localized orbital. The inverse of this small difference tends to be a noteworthy singularity (e.g., $1/\chi$ with $\chi \to 0$).

The above problem arises because of the wavefunctions normalization at the boundary, which has been used in terms of linear-combination-of-atomic-orbital (LCAO)[6] within DFT calculations. The difference between the before and after Kohn-Sham self-consistent field (SCF) calculations, the orbital wavefunction combinations have had the change that away of the normalization condition. Consider Cococcioni's linear response, the perturbation acts through projection $|\psi_i\rangle\alpha_i\langle\psi_i|$ on the *i*th localized orbital with partially filled. However the original orthogonalized wavefunction $\psi_i$ basis set has relaxed to $\psi_i^p$ according to normalization condition during the process of Kohn-Sham equation calculation by SCF, then the change of charge influenced by potential perturbation can be also reflected by $\phi_i^p$ as follows:

$$\sum_i |\psi_i\rangle\langle\psi_i| = \sum_i |\psi_i^p\rangle\langle\psi_i^p| + \sum_i |\phi_i^p\rangle\langle\phi_i^p| = 1 \tag{1}$$

Such effect of wavefunction cannot be detected within the case of fully occupied orbital since it is always that $\sum_i |\psi_i\rangle\langle\psi_i| \approx \sum_i |\psi_i^p\rangle\langle\psi_i^p| = 1$. The *i*th fully occupied orbital occupation deviation, $|\phi_i^p\rangle\langle\phi_i^p|$, is close to zero as the potential perturbation is too small to vary the orthogonal basis sets of $\psi_i$, due to the normalized condition. Therefore the SCF solved difference of orbital occupation is always close to zero and lead the self-consistently determined U parameter to nearly infinitely large.

With identified the main problem, we set out a physical idea to solve the problem instead of directly solving the linear response perturbation on fully-occupied orbitals. In DFT, the Kohn-Sham one-electron eigenvalue is self-consistently solved by minimizing the total energy functional[7], given the condition that the ground state with continuous charge density is defined by $\rho(\bar{r}) = \sum_i n_i |\psi_i(\bar{r})|^2$ and the $n_i$ is a chosen sets follows the Fermi-Dirac distribution which can be fractional. Thus the realistic total energy functional at the ground state with realistic charge density distribution is described as E[$\rho_0$] and with the charge density distribution varied while still follow the Fermi-Dirac distribution law, the new total energy functional has been established with new charge density distribution in the form of $E[\rho] = E[\rho_0] + E_{SE}[\rho] + R[\rho]$. The $E_{SE}$ represents the charge self-energy *functional* of the system induced by the variation of the charge density and the R the orbital relaxation *functional* due to the change of charge occupations. The energy arise from the orbital wavefunction relaxation has been originally systematically discussed by Gopinanthan[8, 9]. But we attempt to elaborate this part with simpler form that easy to understand the idea we illustrated.

The reason we consider only these two effects is that the variation of total energy with respect to the orbital occupation number can be studied with ground- and excited-states in terms of descriptions by Zunger and Freeman[10]. The self-energy has also been primitively discussed as correction into local density formalism by Perdew and Zunger[11], and the self-energy potential can be cancelled by self-exchange-correlation energy on a fully occupied orbital. However, this requires us to make more mathematical efforts to link the view of orbital relaxation effect within ground- or excited states[10].

The Hartree-Fock (HF) method is self-interaction free since the electronic exchange interaction and Coulomb potential are all in the form of $1/|r-r'|$ that offset mutually with exchange integral. But it lacks the electron correlation effect, and therefore there is still a systematic error that vastly overestimated the electronic band gaps. Hybrid functionals always give substantially improved band gaps but closely rely on the empirical choice of the percentage of Hartree-Fock interaction matrix and screening length of short-range electron-electron interactions.

The common exchange correlation functional approximation like local density formalism does not have such above advantages. With help of Hubbard U parameter[12], it is in the presence of HF-like exchange integral in describing the electron repulsion interactions. There have been amount of efforts on using HF theory to described the Hubbard U parameter through determining the on-site electronic Coulomb potential by HF approximations though [13-16], it shows still less presence that follows the Janak theorem as the occupation number of the orbital may not be exactly the same as the density matrix used in HF. Thus in this paper, we attempt to illustrate the above subtle influence with the most simple physical idea with help of local density formalism plus the Hubbard U correction within the framework of orbitals

## II. Physical methodology in orbital theory

Here we approximated that the self-energy and the orbital relaxation with related to the charge occupation are all physical with Fermi-Dirac distribution $n_i$ on ith orbital of charge continuous system, which have been established within diagonalized density matrix in the simplest local density formalism with Hubbard U correction (LDA+U) on a given orbital[17]. To elucidate our physical idea, we choose such simplest LDA+U particular with further simplification of orbital rotational invariant (Anisimov-type)[12, 17], regardless of Hund's-rule spin-exchange coupling (J) and non-collinear effects. The reason for the approximation/assumption is that the boundary conditions are satisfied rigorously and the solutions are expressed by means of eigen function expansions. Therefore, the corresponded eigen function expansion for $E_{SE}$ and R is given $E_{SE}[\rho]=\sum_i n_i \sigma_i$ and $R[\rho]=\sum_j n_j r_j$ in simplification by that physical meaning, which means the modeled on-site self-energy and orbital relaxation can be able to be projected. The symbols of $\sigma_i$ and $r_j$ are therefore the orbital *eigenvalues* of self-energy and orbital relaxation with orbital index i or j.

When the charge density distribution of the system has infinitesimally changed particular to the fully occupied shell, the ground state total energy functional has nearly unchanged ($E[\rho]=E[\rho_0]$) and also obey the relation according to Janak theorem[3] about the orbital eigenvalue $\partial E[\rho]/\partial n_i = \partial E[\rho_0]/\partial n_i = \varepsilon_i$, where the $\varepsilon_i$ is the of *i*th orbital eigenvalue self-consistently (SC) solved by a given exchange-correlation functional used for Kohn-Sham (KS) one-electron equations SC cycles, which is not the real eigenvalue in true world. However, if we pursue the realistic eigenvalues and hope to start from the most simple local density formalism to illustrate this physical picture, we can only provide a loosen idea that current exchange-correlation potential functional (in local density) can only give a minimized residue of the self-energy and orbital relaxation on the *i*th orbital and simultaneously acting as a condition in terms of Janak theorem[3]:

$$\frac{\partial E[\rho]}{\partial n_i} = \varepsilon_i = \varepsilon_{i,real} + \min\left(\frac{\partial E_{SE}[\rho]}{\partial n_i} + \frac{\partial R[\rho]}{\partial n_i}\right) \tag{2}$$

Based on the eigen function expansion at the boundary for wavefunction, the

$$\frac{\partial E_{SE}[\rho]}{\partial n_i} = \sigma_i + \sum_j n_j \frac{\partial \sigma_j}{\partial n_i}, \quad \frac{\partial R[\rho]}{\partial n_i} = r_i + \sum_j n_j \frac{\partial r_j}{\partial n_i} \tag{3}$$

and consider wavefunctions in the system are orthonormal atom-like basis, then we easily treat the above equations with a simple picture in physical that the self-energies of two different orbitals are not interacted mutually. Thus, we got a following generalized condition for fully occupied shell in related to the charge density that follows the Fermi-Dirac distribution and described under the frame work of local density formalism:

$$\sigma_i + r_i \to \min, . \tag{4}$$

This formula guarantees the Janak theorem remains the validity in the true ground state by the local density functional calculations. The Eq. (4) means the under the local density formalism we can choose a suitable sets $n_i$ to simultaneously minimize both the eigenvalues of self-energy and orbital polarization (relaxation energy) of the charge variation for the same orbitals (i=j). *This means in the ideal exchange-correlation functional: the summed contribution of self-energy and orbital relaxation induced energy shift equals to zero.* In addition, the Eq. (4) can be solved by wavefunction projection on specific orbital that we focus, which is shown in the Kohn-Sham (KS) equation of the *i*th orbital from the *I*th atomic site of the system

$$E[n_I]\Big|_{initial}^{KS-SCF} = \min[E[\rho] - \sum_{I,i \text{ or } j}(n_{I,i}\sigma_{I,i} + n_{I,j}r_{I,j})]\Big|_{initial}^{KS-SCF}, \tag{5}$$

where the $n_I$ is the perturbed (relaxed) occupation number of the *I*th atomic site when the residue of $\sigma_i+r_i$ come out. As simplification of the model, we do not consider the orbital components of the *I*th site.

$$E[n_I]\Big|_{initial}^{KS-SCF} = \min[E[\rho] - \sum_I(n_I\sigma_I + n_I r_I)]\Big|_{initial}^{KS-SCF}, \tag{5a}$$

where the $\sigma_I = U_I + \alpha_I$ and $U_I$ is thus on-site (i.e. the *I*th atomic site) self-energy potential induced by the localized orbital due to the repulsive energy of collective Coulomb contributions from fully filled electrons. The $\alpha_I$ is the first-order perturbation of the eigenvalue of the on-site self-energy $\sigma_I$. While $r_I = (1eV) \cdot \delta n_I$ is defined for the change of occupation from $n_I$ to $(n_I + \delta n_I)$ which induced the system initial orbital polarization (while the orbital wavefunctions are still remained the unperturbed) as equivalent relaxation energy, when the system wavefunctions are under the LDA+$U_I$ local environment. As required by physics in Eq. (1), we established in forms of correlation between perturbed occupation number to model the $|\phi_i^p\rangle\langle\phi_i^p|$ by Maclaurin series expansion (a simplified form of Taylor expansion) presented as follow:

$$n_i(\alpha_i) = n_i(0) + \frac{\partial n_i}{\partial \alpha_i} \cdot \alpha_i + o(\frac{\partial n_i}{\partial \alpha_i} \cdot \alpha_i) \cong n_i(0) + (-\frac{\alpha_i}{U}) \tag{6}$$

The above term $\delta n_i$ is therefore described by $\delta n_i = n_i(\alpha_i) - n_i(0) = (-\frac{\alpha_i}{U})$, playing as Lagrange multiplier. And the one-electron effective self-energy potential and relaxation potential approximated with quadratic for derived by the self-energy and orbital relaxation according the condition by Eq. (4):

$$U_I + (U_{SE} - U_R)_I = -\frac{\partial^2 E[n_I]}{\partial n_I^2}\bigg|_{initial}^{KS-SCF}$$
$$= (\frac{\partial \sigma_I}{\partial n_{I,\sigma}} + \frac{\partial r_I}{\partial n_{I,r}})\bigg|_{initial}^{KS-SCF} = U_I + \Delta_I \quad (7)$$

The above operation in Eq. (7) is in fact unchanged at the boundary of wavefunction regardless of whether partially or fully occupied shells before the an *ab-initio* calculation. Therefore, there does a competition (shown as $\Delta_I = (U_{SE} - U_R)_I$ in above Eq. (7)) between two different types of on-site Hubbard-type[1] one-electron potentials (*Note*: It has the same type of interactions but not the same as the Hubbard parameter U.) yielded by the gradient of self-energy and orbital relaxation with respect to an orbital occupation. The Eq. (1) follows the variation principle naturally to yield the Eq. (7) within DFT framework. By solving the derivation of Eq. (1) (i.e. Eq. (7)), we can find out a common set of $n_i$ for simultaneously fulfilling the condition of Eq. (1) by LDA+U self-consistent cycle. That is to say, if $\Delta_I = 0$, we can find out the common set of $n_i$ for simultaneously cancel both $U_{SE}$ and $U_R$ so as to minimize the $\sigma_I$ and $r_I$; if $|\Delta_I| > 0$, this means we cannot find out the unique common set of $n_i$ for both counteraction of $U_{SE}$ and $U_R$. This analysis on the case of $\Delta_I = 0$ recalls the consistency with theoretical framework (Eq. 29 in Ref. [11]) given by Perdew and Zunger[11]. The corresponded $U_I$ is thus the self-consistently determined self-energy potential for consideration of LDA+$U_{SCF}$ calculations, which originally reflected as a Lagrange-multiplier type. But the $U_I$ is the degenerated value, because we only take the degenerated orbital into account.

Now a necessary treatment has been considered for taking the components of the orbital into account, as shown in following Eq. (7a).

$$m^{(-1)^n}U_I + (U_{SE} - U_R)_I = -\frac{\partial^2 E[n_I]}{\partial n_I^2}\bigg|_{initial}^{KS-SCF}$$
$$= (\frac{\partial \sigma_I}{\partial n_{I,\sigma}} + \frac{\partial r_I}{\partial n_{I,r}})\bigg|_{initial}^{KS-SCF} = m^{(-1)^n}U_I + \Delta_I \quad (7a)$$

The $m^{(-1)^n}$ is the effective orbital angular momentum degeneracy correcting coefficient, and *m* is related to the spin-orbit splitting (or j-j coupling) as observed in the core level x-ray photoelectron spectroscopy (XPS). The j=l+s where l is the orbital angular momentum quantum number and s is the spin angular momentum number). Thus, the relationship $m = (1/2)\sum j$ denotes the fully occupied and partially occupied shell corresponding with n=0 and 1 respectively. Since we ignore the orbital component to approximate our model in Eq. (5a) so as to obtain the condition of Eq. (7) with simplified model of orbital relaxation from Eq. (6). Therefore the necessary treatment will be applied at the end to restore the detail components of the focus orbital.

We can also use the linear response method[2, 18] to obtain Eq. (7). Combining Eq. (6) and Eq. (7), we proposed generalized condition for orbital potential perturbation that self-consistently determined the U with Janak theorem maintained:

$$\left(\frac{\partial E}{\partial n_i(a)}\right) - \left(\frac{\partial E}{\partial n_i(b)}\right) = \frac{\delta}{\delta n}\left[(U + \alpha_i(a)) - \left(\frac{\alpha_i(b)}{U}\right)\right] \quad (8)$$
$$= \delta(\alpha_a - \alpha_b)/\delta n = 0$$

The Eq. (8) represents two different linear response processes with two different Lagrange potential perturbations respectively[19]. There is no dependence on empirical data from Eq. (1) to Eq. (8) and only relies on physical meaning of Eq. (1).

With help of orbital potential shift by $U_I$, the boundary of wavefunction for the fully occupied orbital has correspondingly modified as similar form as the partially occupied orbital, with exactly the same absolute orbital eigenvalue (Eq. (2)) and the inter-level of neighboring orbitals ($\partial \varepsilon_i / \partial n_i$). Thus, the output U is $U_{SCF}$ if Eq. (7) is fulfilled with $U_{in}$ for U. Kulik et al have well presented the preliminary relationship of $U_{SCF}$ and $U_{in}$ in LDA+U calculations[18]. By that we rebuilt the formulation of linear response acquired by Eq. (7) and referred to the work of Cococcioni et al[2] and Kulik et al [18] is:

$$U_{out1} = -\left(\frac{\partial \alpha_I}{\partial q(a)} - \frac{\partial \alpha_I^{KS}}{\partial q(a)_{KS}}\right) = -\left(\frac{\partial \alpha_a}{\partial q(a)} - \frac{\partial \alpha_a^{KS}}{\partial q(a)_{KS}}\right) \cdot \left(\frac{U_{in}}{a_0}\right)$$
$$= \left(\frac{U_{in}}{a_0}\right)\left(U_{scf1}(U_{in}) - \frac{U_{in}}{m}\right) \quad (9)$$

$$U_{out2} = -\left(\frac{\partial \alpha_I}{\partial q(b)} - \frac{\partial \alpha_I^{KS}}{\partial q(b)_{KS}}\right) = -\left(\frac{\partial \alpha_b}{\partial q(a)} - \frac{\partial \alpha_b^{KS}}{\partial q(a)_{KS}}\right)$$
$$= U_{scf2}(U_{in}) = \left(\frac{a_0}{U_{in}}\right) \cdot U_{scf1}(U_{in}) = \left(\frac{a_0}{U_{in}}\right) \cdot U_{scf}(U_{in}) \quad (10)$$

Based on Eq. (7), the following condition is determined:

$$|\Delta| = |U_{out2} - U_{out1}|_{U_{in}} = \left|U_{scf}\left(\frac{a_0}{U_{in}} - \frac{U_{in}}{a_0}\right) + \left(\frac{U_{in}}{a_0}\right)\left(\frac{U_{in}}{m}\right)\right|_{U_{in}} \geq 0 \quad (11)$$

Janak theorem in fact returns the angular momentum degenerated eigenvalues, the self-consistent solved orbital eigenvalue is actually degenerated by orbital angular momentum. This means the m of the Eq. (12) is decided by *post-hoc*. This leads us to have effective degeneracies, $m_1$ and $m_2$ to decide, which presented as only parameter unknown in the work of Kulik et al[18]. In addition, these effective orbital degeneracies are defined differently from the results of Kulik et al[18]. However, this would not make Eq. (12) be semi-empirical. Then, by considering two different effective degeneracies ($m_1$ and $m_2$) for the fully and partially filled shells, respectively, the two different cases above are summarized to self-consistently determine and lock the input $U_{in}$ to obtain $U_{scf}$ in the DFT+U calculations:

$$U_{scf} = \begin{cases} \dfrac{\left(\dfrac{U_{in}}{m_1}\right)}{\left(1-\left(\dfrac{a_0}{U_{in}}\right)^2\right)} & (occ=1, crossover) \quad \left(m_1 = \dfrac{1}{l_{max}}\right) \\ \\ \dfrac{\left(\dfrac{2U_{in}}{m_2}\right)}{\left(1+\left(\dfrac{a_0}{U_{in}}\right)^2\right)} + \Delta & (non-crossover) \quad \left(m_2 = \left[(2l+1)-\dfrac{occ}{2}\right]\right) \end{cases} \quad (12)$$

For a system of many-body problem, the orbital with different occupations of the particle inevitably induce systematic error by KS equation used in DFT calculation, since the effective coulomb potential cannot uniquely described the detail interactions of various orbitals with different occupation numbers. Thus, our model is advantages to deal with large scaled solid material for efficiencies but maybe still lack detailed orbital-by-orbital interaction information.

### III. Validation test results and discussion

Here, we provide some interesting comparisons with experimental data, with respect to the DFT+U calculations. Figure 1 illustrates how the self-energy to be counteracted in fully occupied shell of cations and how self-energy influences the partially filled shells. The crossover point of the Figure 1 shows the validation of the Eq(8)=0 based on condition of Eq (1). Figure 2 demonstrates the on-site Coulomb potential as U parameters for ZnO, GaAs, $Cu_2O$ and CuO. As summary in Table 1, we benchmark all of the fully occupied shell materials based annihilated self-energy. As is well known, the ZnO, CdO and GaN has relative high 3d levels led to underestimation of p-d couplings between 2p (or 3p for S element and P elements) and 3d/4d orbitals, the band gap bewteen highest occupied and lowest unoccupied bands is consequently underestimated. Even the 3d level of $Ga^{3+}$ in GaN is very deep around -22 eV below the VBM (0 eV), but it still plays an important role on giving contributions of the hybridization bonding with 2p orbitals of N, influencing the lattice parameters and more importantly to the band gap. The band gap of GaAs is 0.2 eV higher than the experimental band gap at 0 K. This arises because the 4p level of $As^{3-}$ in GaAs is less localized to give evident and direct influence on the band gap compared to the case in GaN. Even though the 3d levels show the core states in ZnS, CdS, GaN, GaP, and GaAs, but the p-d repulsive potential still contributes the band gaps, and it covers the 3d semicore states and the top of valence bands.

We use mean relative error (MRE) to give a clear quantitative comparison with other method reported. From Table 1, the LDA+$U_{SCF}$ provide relaxed lattice relaxation results with MRE of 0.76%. As we investigated from the plain PBE functional[20] (a gradient approximation based on LDA) is found to be 1.1% in general[21]. Compared to the results from hybrid-functional calculations, the MRE is relatively satisfied as it is found to be 2.3% for screened exchange and 0.2% for HSE[21]. Consider the electronic band gaps, our overall MRE for LDA+$U_{SCF}$ is 3.89% for metal oxides with fully occupied orbital, while 38% for plain PBE method and 7.4% for hybrid functional method for all semiconductors and insulators as reported[21, 22]. We see that the maximum displacement occurs in the results of GaAs with 0.24 eV larger than the experimental data. This arises because the 3d states of Ga move even deeper than the other gallium picnitides (-22.8 eV, VBM=0eV), it is not certain to the interplay of the weak p-d repulsion effect and minimum band gap for GaAs after our U corrections.

We further choose ZnO for example to illustrate the evident errors induced by inappropriate chosen of the U parameters, as well as the comparison between DFT+U and other methods (Table 2). The fraction of 0.375 shown in Table 2 is the percentage of HF interaction matrix using in HSE hybrid-functional from VASP package. We have shown a competitive accuracy compared to the other DFT calculations from other groups [24-28]. The plain PBE calculations based on non-linear core corrected (NLCC) norm-conserving pseudopotentials show improved formation enthalpies of ZnO, band gaps, and lattice parameters, but underestimated the 3d levels, compared to the plain PBE calculation with conventional norm-conserving pseudopotentials[21, 23]. Moreover, the contribution of the NLCC of the norm-conserving pseudopotential method has been introduced in the previous work[24]. The PBE+$U_{scf}$ ($U_{scf}$ denotes the U parameters are self-consistently determined by our method mentioned above and illustrated in Table 1) based on NLCC norm-conserving pseudopotentials shows substantially improved results of formation enthalpy, band gap, lattice parameters, and 3d levels, as well as the formation energy oxygen vacancy in the low oxygen chemical potential limit (O-poor), excellently corresponds to the experimental data and consistent with the other calculation results [24-28].

We recall the previous work done by Ma et al who has empirically tuned U parameters for 3d orbitals of Zn and 2p orbitals of O respectively, and shown a correlation effect[25] in terms of band structure, lattice geometry, and native defect levels. With consideration of both U parameters on $3d^{10}$ of $Zn^{2+}$ and $2p^6$ of $O^{2-}$, the lattice constant, bulk modulus, and other bulk properties have surprisingly close to the experimental data, when the band structure has been tuned to the experimental data, this effect is even valid to the oxygen vacancy formation energy. Such effect seems to occur regardless the local atomic coordination as they investigated among zincblend, rocksalt, and wurtzite phases of ZnO, which hints an intrinsic feature. We applied our physical model on the 2p orbital of O sites and found well consistent with their results (6.5 eV in ours and 7.0 eV in Ma et al[25].), and our self-consistently calculated Ud is also showing similar magnitude compared to the Ud=10 eV discovered by manually tuning[25]. With combination of Ud and Up for Zn and O respectively, we observed the error in both band structure and relaxed lattice constant have been vastly minimized compared to experiments, which confirm the successfulness on fully occupied orbital like ZnO by local density formalism with Hubbard U parameter correction. The displacement of values may due to the choice of the different pseudopotentials of atoms in the solids. This comparison also confirmed the interplay of p-d orbital entanglement.

Figure 3 shows the band structures of the wurtzite ZnO and zincblend ZnS as a comparison about how the $3d^{10}$ level moves for Zn in related oxide and sulfide. The 3d level for ZnO is broader than the ZnS and stays as high as -7.5 eV (VBM: 0eV) but -10.7 eV in ZnS. The 3d level of ZnO has remarkably close to the experiment by high-resolution and resonant angle-resolved photoemission spectroscopy (ARPES) with reported -7.5 eV as well[26]. In addition, the s levels of S sites in ZnS move even higher than the 3d level of Zn, presenting a trend of hybridization of s-p orbitals of S. The s band in ZnS has overlap with 3d band at the Γ of the first Brillouin-Zone (BZ). It shows a relatively strong p-d entanglement in ZnO compared to ZnS as the 3d level move further deep in the latter material with 3.2 eV.

Figure 4 shows the band structures of CdO and CdS within zincblend lattice. The CdO has an indirect band gap with 1.38 eV (direct: 2.46 eV) while the CdS has a direct band gap of 2.65 eV. The 4d band for CdO starts from -7.8 eV with slightly deeper (0.3 eV) than the case in ZnO. The 4d band in CdS shows 4.5 eV deeper than the band in CdO.

Figure 5 shows the band structures gallium picnitides (GaN, GaP, and GaAs) in zincblend phase. The band gap calculated by PBE is generally underestimated and about ~1.5 eV for GaN. We see that the band gap has increased to about 3.5 eV by the Hubbard U corrections. The valence bands of them have s-p hybridized feature shown in the projected DOS. The Figure 5 also shows the s band minimum of N in GaN is relatively flat near the Γ, while a slightly distorted in GaAs. This may due to the strong repulsion effect by $3d^{10}$ of Ga as the $3d^{10}$ level move lower than the s band in GaAs. Experiment reported the 3d level of Ga is from -18 eV to -22 eV[27], we shows a consistent results based on our calculation model, while the hybrid-functional gives -16 eV with evident difference[21].

Figure 6 shows the band structures of $Cu_2O$ and CuO (AFM). The Cu has $3d^{10}$ configuration in $Cu_2O$ while $3d^9$ for CuO. It turns out a difficult task to obtain accurate band structures as reviewed from screened-exchange, the HSE[28] or GW methods[29], even as early as 1980s using the pseudopotential-based tight-binding view [30]. The most difficult step is to deal with correct positioning the 3d orbital levels of $Cu_2O$, which is higher than the 2p bands of O and totally different from other semiconductor metal oxides like ZnO (Figure 3). The valence band of CuO consists of obvious overlap between 3d filled states of Cu and 2p of O, while the conduction band minimum is contributed by the un-occupied state of 3d orbital of Cu.

Figure 7 shows the band structures of $Lu_2O_3$ as our extension test on the fully occupied 4f orbitals, expecting our model still valid. The results show positive support as the band gaps of different phases are almost 5.5 eV close to the experimental reported by Prokofiev et al[31]. The fully filled $4f^{14}$ level started from about -6.5 eV below the VBM (0 eV). The valence band width is about 4~5 eV close to results by other group[32], but with improved electronic direct band gaps calculations.

Finally, Figure 8 (a) concentrates the procedure of our model and elucidates the validation for the partially occupation number of electron. Lany and Zunger have considered the oxygen localized hole state problem [33, 34] in wide band gap semiconductor oxides and proposed a method resembling the Koopmans theorem[34]. Thus, we determine the relationship between the system electron energy and its integer occupation number. As illustrated by Figure 8 (b), the correct description of localized states depends on $d^2E/dn^2$[35-37]. The second derivative of E is a concave feature for HF theory regarding the continuous occupation number, which is $d^2E/dn^2<0$, whereas it is a convex feature for LDA/GGA, $d^2E/dn^2>0$. However, the correct behavior is actually linear[35-37], which means that $d^2E/dn^2 =0$. Therefore, the proposed method is described as

$$\left| U_{out2} - U_{out1} \right|_{U_{in}} = \frac{d^2 E}{dn^2} = 0 \tag{13}$$

The E in Eq(13) is not the system total energy. It is actually the electronic energy with integer occupation number. The work we elucidated here is an attempt to illustrate that the local density formalism with Hubbard U correction does have physical meaning in different angles, which has also emphasized in pioneer's work of Cococcioni et al [2] and Hulik et al[18]. As is well known, U is not a fitting parameter, but an intrinsic electronic Coulomb response property. As introduced by Cococcioni and de Gironcoli[2], we agree that the Hubbard U parameter represents the spurious curvature of the LDA/GGA energy functional as a function of occupations, and LDA/GGA+U does exhibit the piecewise-linear behavior of the exact ground-state energy. Thus, we aim to use a simple model to support the work [2, 18] that U in DFT+U method has physical meaning.

As we known, there have been considerable efforts to correct the self-energy error if they treat it only a spurious error part within their calculation method/functional. Based on our work here, however, it can be endowed with physical meaning solved in the orbitals. The simplest method is to use the DFT+U method, which adds an on-site repulsive potential U for the localized d and f orbitals of solids to cancel the self-energy[12, 17]. Even this simple method, however, the choice of U parameter for correction is mostly semi-empirical to match the experimental reported data, but still led to error of the band structure calculations especially in the defect state calculations. In fact, the defect state levels in the band structure arise from the host in the Brillouin zone and correlated with the excited states of the band structure. The method of calculation that closely describes the band structure may not be sufficient enough to correctly reflect the defect states, in particular to the deep levels. Therefore, we need to be careful to determine the U parameters for further defect state studies.

## IV. Summary

In summary, we pointed out that the strong boundary normalized condition of wavefunction for fully occupied semicore 3d orbitals leads the linear response DFT+U on such metal oxide to have a difficulty in Hubbard U determination. With identified the main problem, we treated the orbital self-energy and orbital relaxation as projected components of on-site orbital eigenvalues with respective orbital occupation number that follows the Fermi-Dirac distribution. By self-consistently solving the second partial deviation of total energy based on the most simple local density formalism with Hubbard U correction, we found the local density exchange-correlation potential functional can only give a minimized residue of the self-energy and orbital relaxation on the focus orbital if the Janak theorem maintained. Such residue turns to well counteracted in the fully occupied orbitals and non-zero rigid difference for the partially occupied orbitals. With maintaining the validation of Janak theorem on localized orbitals, the self-consistent cycle by local density functional with Hubbard U correction cannot find out a set of orbital occupation that simultaneously offsets the orbital self-energy and relaxations in the empty and partially filled shell, but returns a unique set of the occupation for fully occupied shell. Using this method, we applied the self-consistently determined on-site Coulomb potential to the Hubbard U parameters of the fully occupied d orbitals of cations and p orbitals of anions in a compound. The band gap calculations on fully occupied orbital based compounds are thus improved and the relaxed lattices are also shown based on minimization of the self-energy error, which shows a possible route for accurate excited state studies. Thus we hope, by using this simple model, we can support the work [2, 18] that U in DFT+U method has physical meaning with more evidence. We expect this method provides electronic structures of the eigen-bulk properties and satisfies the native defect levels of bulk or low-dimensional structures. We also expect this method will accelerate the pace of electronics research and development (R&D) by engineering new generation of materials that form or are synthesized in extreme physical or chemical environments.

**Acknowledgement**

The author gratefully acknowledges the support of natural science foundation of China (NSFC) for Youth Scientist (Grant No. NSFC 11504309).

**Table 1.** Comparison of the calculated PBE+U(scf) on binary filled shell compounds to the experimental results. The experimental band gap are referred to the ones from 0K. (Refs. [21, 28, 38-44])

| Compound | Ud (scf) (eV) | Up (scf) (eV) | Lattice (Å) | EXP. Lattice (Å) | Eg (eV) | EXP. Eg (eV) | $E_{d/f}$ (eV) | EXP. $E_{d/f}$ (eV) |
|---|---|---|---|---|---|---|---|---|
| ZnO (hex) | 11.238 | 6.502 | 3.248/5.216 | 3.249/5.205 | 3.441 | 3.44 | -7.5 | -7.3 |
| ZnS (cub) | 13.083 | 5.174 | 5.470 | 5.409 | 3.846 | 3.8 | -10.7 | |
| CdO (cub) | 10.102 | 6.862 | 4.704 | 4.695 | 1.379/2.459 | 2.3(dir) | -7.8 | |
| CdS (cub) | 13.210 | 5.061 | 5.893 | 5.832 | 2.652 | 2.4 | -12.3 | |
| GaN (cub) | 17.878 | 6.866 | 4.552 | 4.540 | 3.534 | 3.504 | -21.2 | -22.2 |
| GaP (cub) | 17.513 | 4.802 | 5.567 | 5.450 | 2.491 | 2.32 | -22.6 | |
| GaAs (cub) | 16.205 | 4.779 | 5.717 | 5.653 | 1.755 | 1.519 | -22.8 | -22 |
| $Cu_2O$ | 6.850 | 13.122 | 4.276 | 4.270 | 2.159 | 2.17 | -4.5 to 0 | |
| CuO (AFM) | 6.361 | 4.322 | 4.558/3.644/5.207 | 4.653/3.410/5.108 | 1.681 | 1.7 | -9.5 to -2 | |
| $Lu_2O_3$ (hex) | 13.976 | 5.467 | 3.604/5.812 | | 5.584 | 5.5 | -6.8 | |
| $Lu_2O_3$ (cub) | 12.307 | 5.497 | 10.472 | 10.391 | 5.528 | 5.5 | -6.0 | -5.8 |

**Table 2.** Benchmark of the present work with other calculation methods in different packages.

| Benchmark | | a (Å) | c (Å) | Eg (eV) | $E_{3d}$ (eV) | $\Delta H_f$ (eV) | Vo (O-poor) | Refs |
|---|---|---|---|---|---|---|---|---|
| VASP | PBE_PAW_U | 3.148 | 5.074 | 1.51 | -6.0 | -3.50 | 3.72 | [45] |
| | PAW_HSE | 3.261 | 5.225 | 2.49 | -5.8 | -3.01 | 0.96 | [46] |
| | PAW_HSE_(0.375) | 3.249 | 5.196 | 3.43 | -6.4 | -3.13 | 1.01 | [46] |
| | PAW_GW | | | 3.34 | | | | [29] |
| CASTEP | PBE+NoNLCC | 3.286 | 5.299 | 0.90 | -4.8 | -2.93 | | [22, 23] |
| | PBE+NLCC | 3.278 | 5.301 | 1.04 | -4.0 | -3.20 | | [this work] |
| | sX | 3.267 | 5.245 | 3.41 | -7.0 | -3.31 | 0.85 | [22, 23] |
| | PBE+U(scf)+NLCC | 3.248 | 5.216 | 3.441 | -7.53 | -3.70 | 0.963 | [this work] |
| Exp. | Exp. (1) | 3.242 | 5.188 | 3.44 | -7.5 | -3.63 | | [47, 48] |
| | Exp. (2) | 3.249 | 5.205 | 3.44 | -7.3 | | | [38, 39] |

**Figure 1**

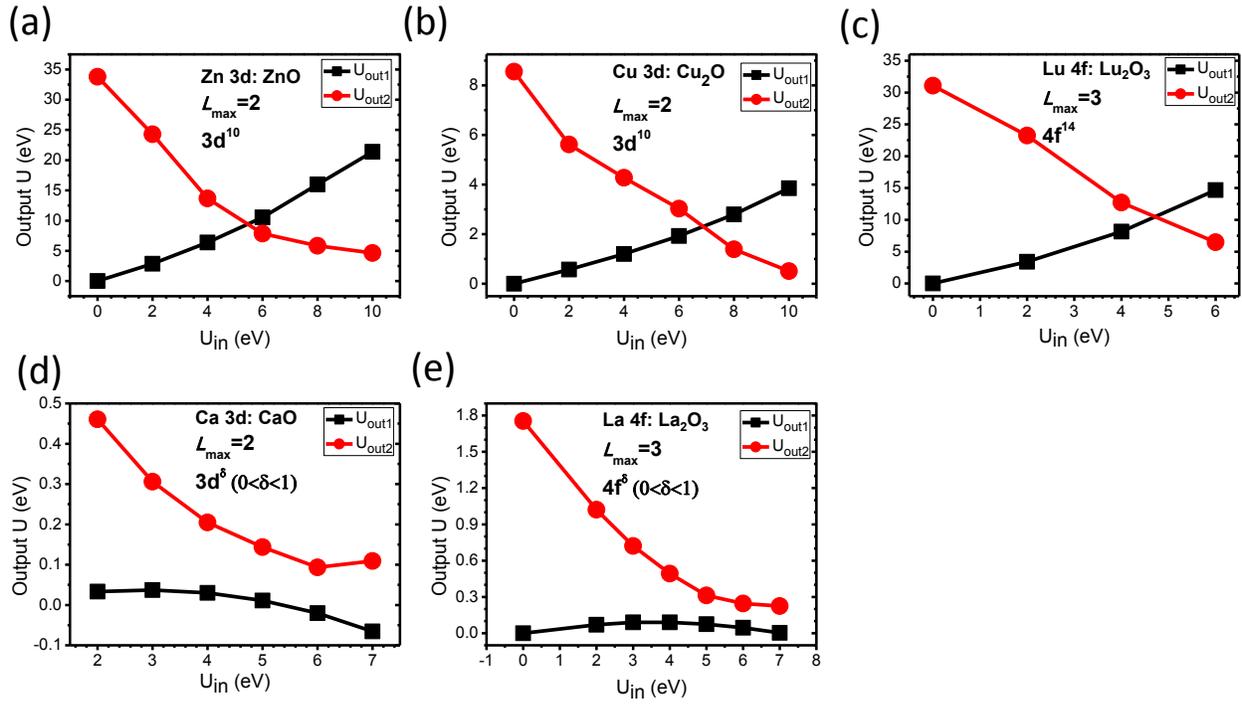

**Figure 1**. Obtained $U_{out1}$ and $U_{out2}$ for fully occupied orbitals from (a) ZnO, (b) Cu$_2$O and (c) Lu$_2$O$_3$. The cross-over feature denotes the $|U_{out1}-U_{out2}|=0$. This shows the contrast for (d) CaO and (e) La$_2$O$_3$ with partially occupied orbitals, without cross-over.

**Figure 2**

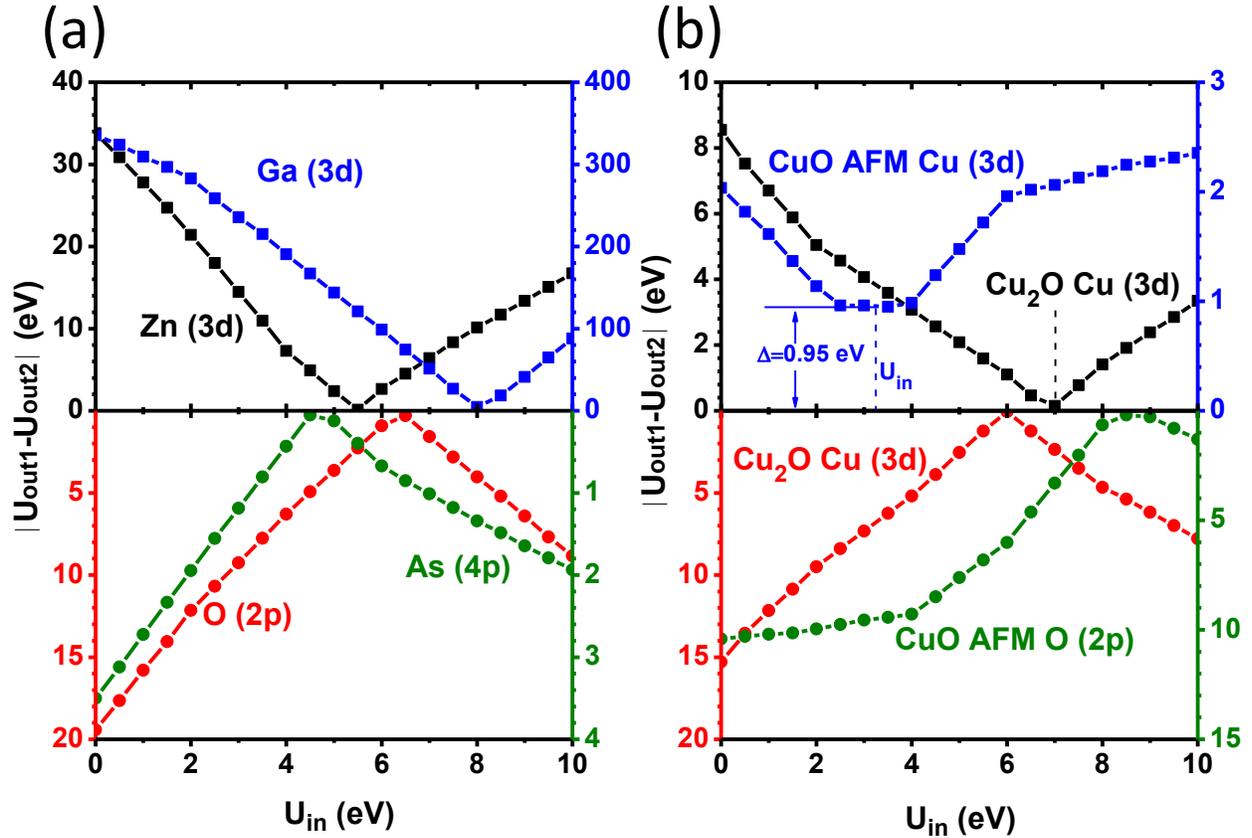

**Figure 2**. The $|U_{out1}-U_{out2}|$ vs $U_{in}$ behaviors of bulk wurtzite ZnO, zincblend GaAs, bulk CuO in AFM phase, and bulk $Cu_2O$ structures with d and p localized electronic orbitals. (AFM: anti-ferromagnetic).

**Figure 3.**

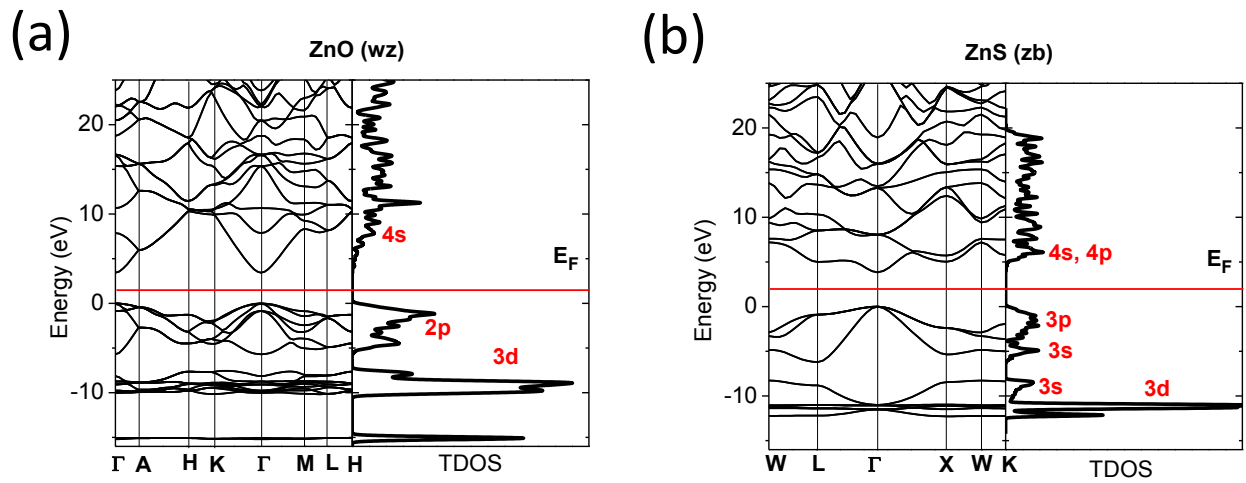

**Figure 3.** Band structures of ZnO in wurtzite phase (a) and ZnS in zincblend phase (b).

**Figure 4.**

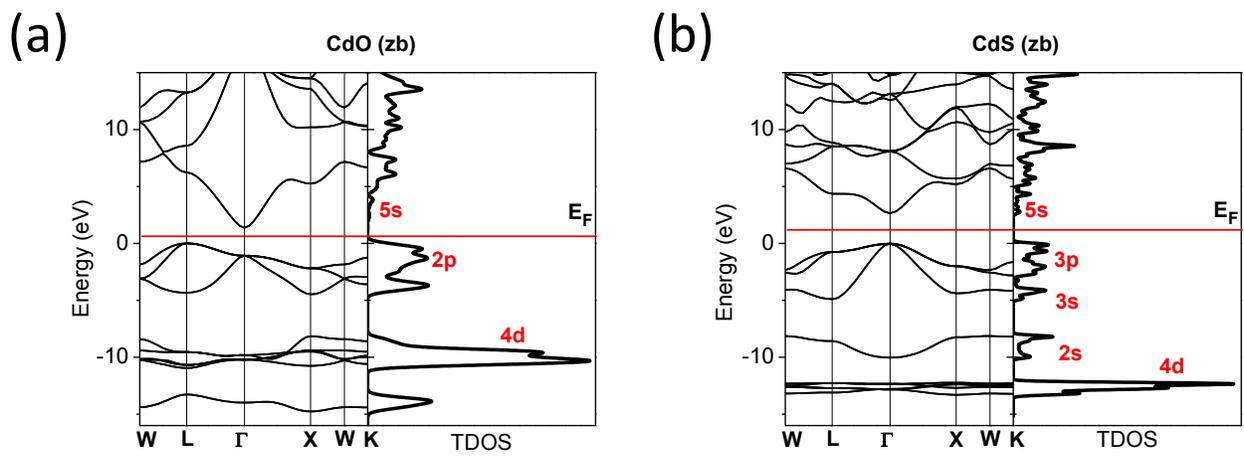

**Figure 4.** Band structures of CdO (a) and CdS in zincblend phase (b).

**Figure 5.**

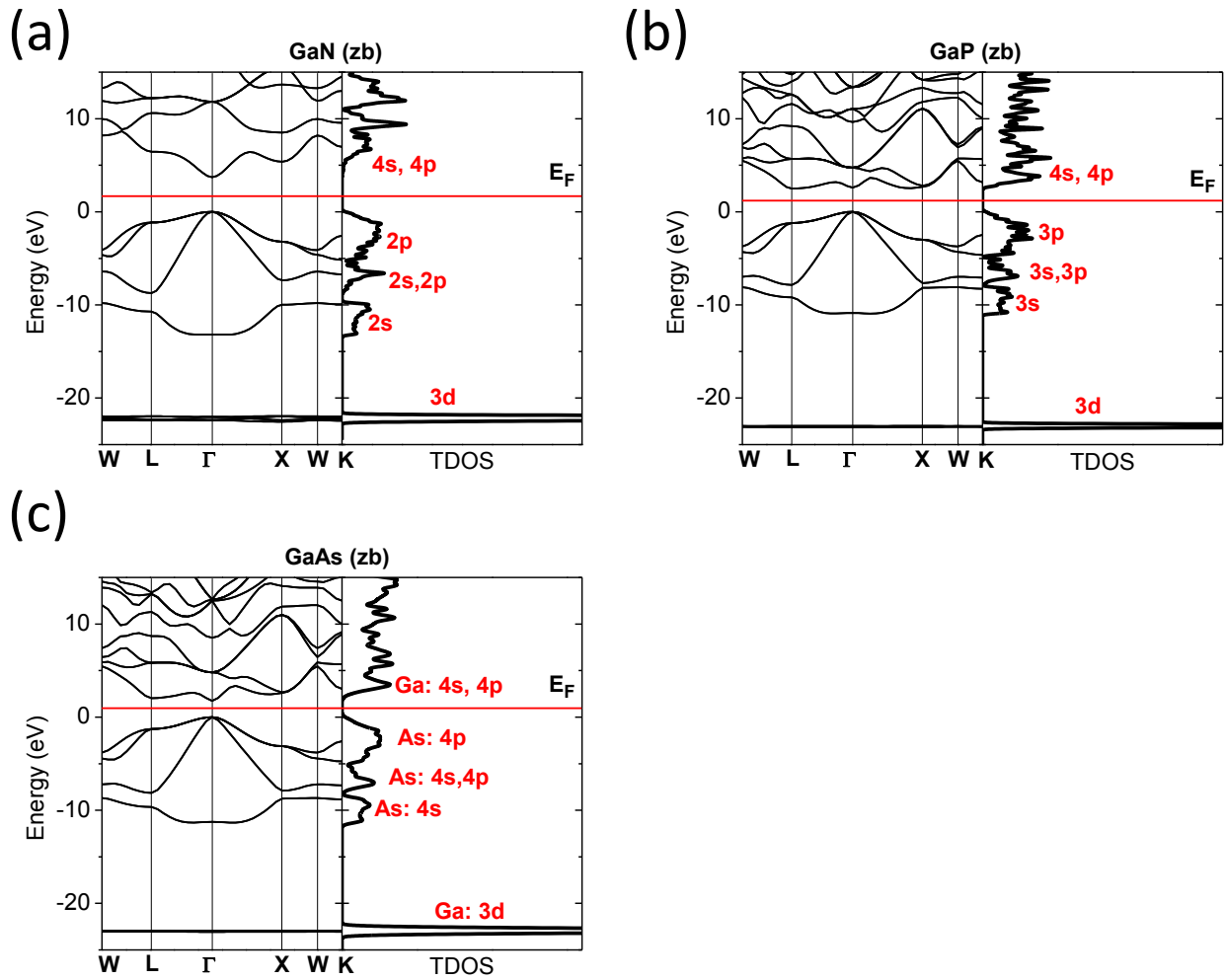

**Figure 5.** Band structures of GaN (a), GaP (b) and GaAs (c) in zincblend phase.

**Figure 6.**

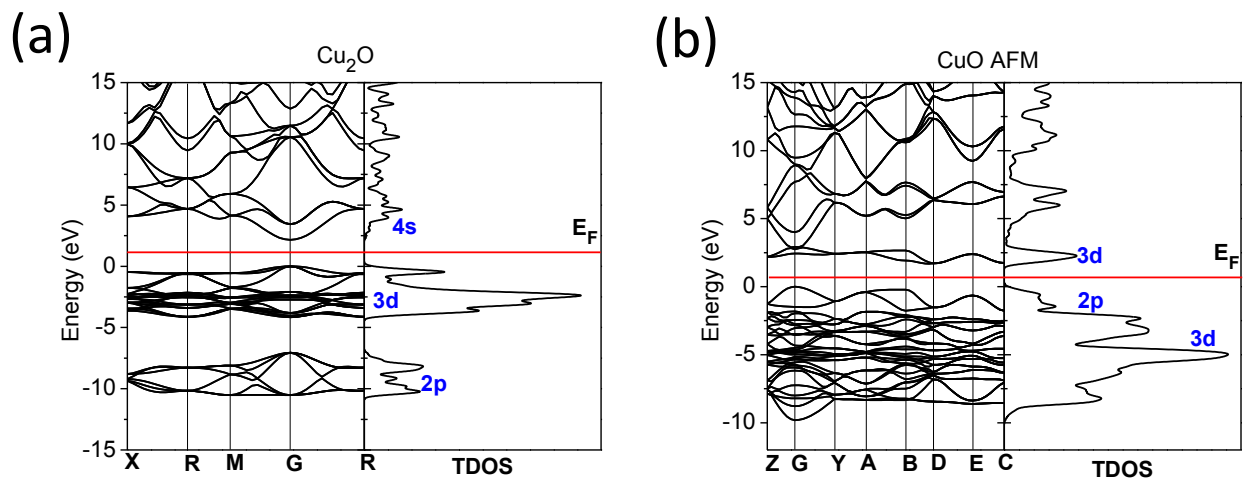

**Figure 6.** Band structures of $Cu_2O$ (a) and CuO in antiferromagnetic (AFM) phase (b).

**Figure 7.**

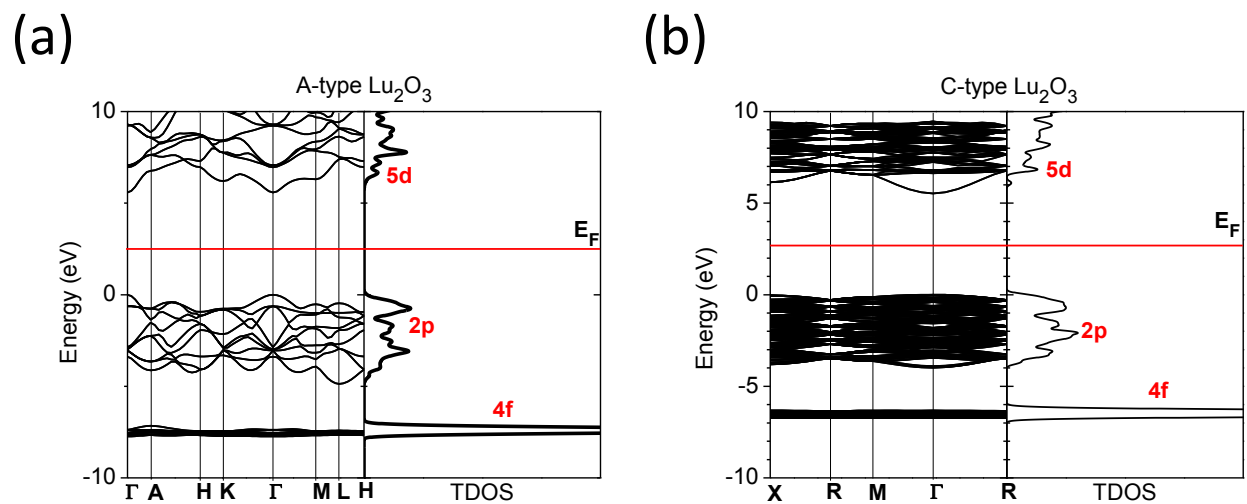

**Figure 7.** Band structures of $Lu_2O_3$ in hexagonal (A-type) (a) and cubic lattices (C-type) (b).

**Figure 8.**

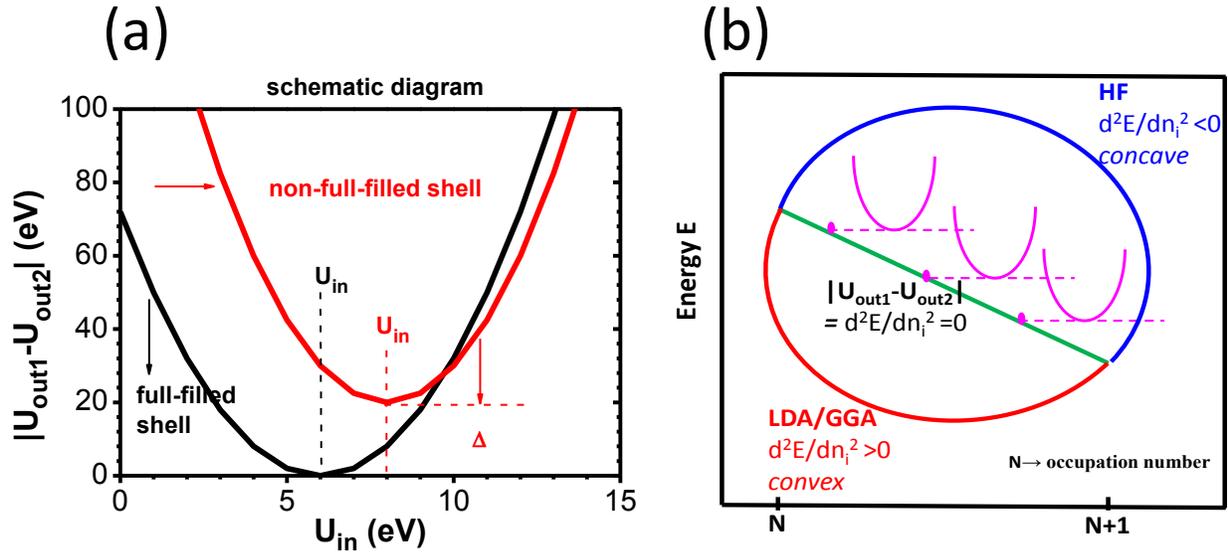

**Figure 8.** (a) The $|U_{out1}-U_{out2}|$ vs $U_{in}$ behaviors varied by full-filled and non-full-filled shells. (b) Electronic energy vs integer/fractional occupation numbers with different theoretical models for dealing with many-body calculations.